\documentclass[11pt]{article}

\usepackage[margin=1in]{geometry}

\usepackage{mathtools, amssymb, array, marvosym}
    \def\RR{\mathbb R}
    
    \def\geq{\geqslant}
    \def\leq{\leqslant}
    \def\ee{\varepsilon}
    \def\BBB{\mathcal B}
    \def\CCC{\mathcal C}
    \def\DDD{\mathcal D}
    \def\EEE{\mathcal E}
    \def\OOO{\mathcal O}
    \DeclareMathOperator\BSC{BSC}
    \newtheorem{theorem}{Theorem}
    \DeclareMathOperator\poly{poly}
    \DeclareMathOperator\Prob{Prob}
    \DeclareMathOperator\Roll{Roll}
    
    \DeclareMathOperator\Capacity{Capacity}
    \def\bma#1{\begin{bmatrix}#1\end{bmatrix}}

\usepackage[rgb, dvipsnames, svgnames]{xcolor}
    \definecolorseries{no-y}{hsb}{last}[hsb]{.2,1,1}[hsb]{1.1,1,1}

\usepackage{tikz-cd, caption, subcaption}
    \usetikzlibrary{decorations.pathreplacing, lindenmayersystems}
    \tikzset{every picture/.style={line cap=round, line join=round}}

\usepackage{snaptodo}
    \snaptodoset{margin block/.style={font=\tiny}}

\usepackage[colorlinks, allcolors=blue!50!magenta!50!black]{hyperref}

\usepackage{autonum}

\begin{document}

                                 \title
                    {Capacity-Achieving Gray Codes}
                                    
                                \author
                {Venkatesan Guruswami and Hsin-Po Wang}
                                    
                         \def\day#1\year{\year}
                                    
                               \maketitle

\begin{abstract}
    To ensure differential privacy,
    one can reveal an integer fuzzily in two ways:
    (a) add some Laplace noise to the integer, or
    (b) encode the integer as a binary string and add iid BSC noise.
    The former is simple and natural while
    the latter is flexible and affordable,
    especially when one wants to reveal a sparse vector of integers.
    In this paper, we propose an implementation of (b)
    that achieves the capacity of the
    BSC with positive error exponents.
    Our implementation adds error-correcting functionality to
    Gray codes by mimicking how software updates
    back up the files that are getting updated (``coded Gray code'').
    In contrast, the old implementation of (b) interpolates between
    codewords of a black-box error-correcting code (``Grayed code'').
\end{abstract}

\footnotetext[0\def\thefootnote{\$}]{
    Research supported in part by
    NSF grant CCF-2210823 and a Simons Investigator Award.
    We gratefully acknowledge the hospitality of the Simons Institute
    for a dedicated semester on error-correcting codes
    where this work was carried out.
    We extend our deepest gratitude and respect
    to the late Jim Simons (1938--2024)
    for his commitment to investing in mathematics and science.
}

\footnotetext[0\def\thefootnote{\Letter}]{
    Emails: \{venkatg, simple\} @berkeley.edu.
}

\section{Introduction}

    Differential privacy is the art of publishing collective facts
    without leaking any detail of any user.
    A mathematically rigorous way to do so is adding noise to
    an aggregation function that is Lipschitz continuous
    (sometimes of bounded variation) in every argument.
    More concretely, suppose that we are interested in a feature
    $\varphi\colon \{0, 1\}^n \to [m]$ that satisfies
    \[
        |\varphi(u) - \varphi(u')| < 1, \text{ for }
        u \coloneqq (u_1, \dotsc, u_i, \dotsc, u_n) \text{ and }
        u' \coloneqq (u_1, \dotsc, 1{-}u_i, \dotsc, u_n),
    \]
    i.e., changing the data of the $i$th user
    does not change the feature too much.
    Then publishing $\varphi(u) + L$, where $L$ follows
    the Laplace distribution with decay rate $\ee$,
    is $\ee$-differentially private \cite{DMN16}.
    That is,
    \[
        \Prob\{ \varphi(u) + L < t \} \leq \exp(\ee)
        \Prob\{ \varphi(u') + L < t\}                 \label{eq:private}
    \]
    for any number $t \in \RR$, meaning that a data broker
    will have a hard time telling if $u_i$ is $0$ or $1$.

    Publishing $\varphi(u) + L$ is called the \emph{Laplace mechanism}
    \cite{DMN16}.
    It is optimal privacy-wise as
    \eqref{eq:private} assumes equality half of the time.
    But it turns out to be randomness-costly and space-inefficient
    when we have many features $\varphi_1, \dotsc, \varphi_\ell$
    to publish, wherein only $k \ll \ell$ of them are non-zero\footnote{
    For example, $\varphi_i(u)$ could be the number of times
    the $i$th English word was mentioned in a forum archive $u$.
    Most word counts are going to be zero.}
    for a given $x$.
    In this case, the Laplace mechanism will
    add noise to all $\varphi_i(u)$ and then publish all $\ell$ of them.
    For one, this means that we are forced to
    sample Laplace distribution $\ell$ times.
    Even if we can afford that,
    the output will be $\Omega(\ell \log m)$ in size
    ($m$ is an upper bound on the $f$'s)
    while the raw data is only $\OOO(k \log(\ell) \log(m))$.

    A brilliant idea of Lolck and Pagh \cite{LoP24},
    which is a generalization of an earlier work by
    Aumüller, Lebeda, and Pagh \cite{ALP22},
    reduces the space requirement as well as the sampling cost.
    The idea is that, instead of working on the ordered field $\RR$,
    we encode each $\varphi_i(u)$ as a binary string
    $\EEE(\varphi_i(u)) \in \{0, 1\}^{1\times n}$
    and put the bits of $\EEE(\varphi_i(u))$ at $n$
    random places on a tape of length $\Theta(kn)$.
    This is illustrated in Figure~\ref{fig:tape}.
    Note that $\EEE(\varphi_{i_1}(u))$ and $\EEE(\varphi_{i_2}(u))$
    might end up choosing the same random places.
    Such a collision is resolved, fairly, by putting a random bit there.
    We also put a random bit at every empty place.
    These random bits will play the role of the Laplace
    noise---protecting privacy by making precise decoding impossible.

    One problem remains:
    To what extent can we
    translate the binary tape back to real numbers? 
    This motivates the definition of robust Gray codes.

\begin{figure}
    \centering
    \begin{tikzpicture} [y=0.5cm]
        \pgfmathsetseed{8881616}
        \resetcolorseries[5]{no-y}
        \draw [decorate, decoration=brace]
            (0, 0) -- node [above] {$n$ bits} (1.6, 0)
        ;
        \fill [gray!50!white, shift={(2, -3)}]
            (0, 0) -- (1, 3) -- (1, 2) to [bend right=15]
            (8, 2) -- (8, 3) -- 
            (9, 0) -- (8, -3) -- (8, -2) to [bend right=15]
            (1, -2) -- (1, -3) -- (0, 0)
            foreach [count=\x] \X in
            {S,T,R,E,T,C,H,~,O,U,T,~,R,A,N,D,O,M,L,Y} {
                ({(\x+0.5)*8/20}, 0) node [white]
                [xscale=1.3, yscale=2.5+(\x-10.5)^2*0.05] {\bfseries\X}
            }
        ;
        \foreach \y in {1, ..., 5} {
            \tikzset{no-y!![\y]!80!black}
            \draw (0.1, -\y) node [left] {$\EEE(\varphi_{i_\y}(u)) =$};
            \draw (1.9, -\y) node {$\leadsto$};
            \foreach \x in {1, ..., 9} {
                \pgfmathtruncatemacro\r{random(0, 1)}
                \pgfmathtruncatemacro\s{random(0, 5)}
                \draw (\x/6 , -\y) node {$\r$};
                \draw (1.1+\x+\s/6 , -\y) node {$\r$};
                \expandafter
                \ifx\csname\x.\s\endcsname\relax
                    \expandafter\xdef\csname\x.\s\endcsname
                    {\noexpand\color{no-y!![\y]!80!black}\r}
                \else
                    \expandafter\xdef\csname\x.\s\endcsname{c}
                \fi
            }
        }
        \foreach \x in {1, ..., 9} {
            \foreach \s in {0, ..., 5} {
                \def\result{\csname\x.\s\endcsname}
                \expandafter
                \ifx\csname\x.\s\endcsname\relax
                    \xdef\result{e}
                \fi
                \draw
                    (1.1+\x+\s/6, -6.5) node [align=center] {
                        \result\kern-0.5em
                        \raisebox{0.7cm}
                        {\rotatebox{-90}{$\leadsto$}}
                        \kern-0.2em
                    }
                ;
            }
        }
        \draw [decorate, decoration=brace]
            (11, -8) -- node [below]
            {tape is $\Theta(kn)$ bits} (2, -8)
        ;
    \end{tikzpicture}
    \caption{
        A space-efficient differential privacy mechanism.
        Step 1: encode integers as binary strings.
        Step 2: spread out the bits.
        Step 3: superimpose them on a tape.
        Functions $\varphi_{i_1}(u), \dotsc, \varphi_{i_5}(u)$
        are the ones that are nonzero.
        Labels c and e mean collision and empty, respectively;
        they both will be replaced by random bits.
    }                                                   \label{fig:tape}
\end{figure}

\subsection{Robust Gray Codes}

    A Gray code encodes integers as binary strings
    such that any two consecutive strings differ at exactly one place.
    A popular construction of Gray codes
    is via the ruler sequence \cite[A001511]{oeis}
    \[
        \rho_j \coloneqq
        \text{the greatest number $r$ such that $2^r$ divides $2j$}.
                                                        \label{eq:ruler}
    \]
    The first few terms read
    $1, 2, 1, 3, 1, 2, 1, 4, 1, 2, 1, 3, 1, 2, 1, 5$.
    Then, the $(j + 1)$th string of the $k$-bit reflected Gray code
    is obtained by flipping the $\min(\rho_j, k)$th bit
    of the $j$th string.
    For simplicity, we will write $\min(\rho_j, k)$ as $\rho_j$,
    and so we can write $g^{j+1} = g^j + e^{\rho_j}$
    instead of $g^j + e^{\min(\rho_j, k)}$,
    where $e^r$ is the $r$th standard basis vector of length $k$.
    As an example, when $k = 4$,
    \[\begin{tikzpicture} [y=0.5cm]
        \resetcolorseries[4]{no-y}
        \foreach [count=\y] \r in {1,2,1,3,1,2,1,4} {
            \draw [line width=0.3cm, {no-y!![\r]!50!white}]
                (\r*0.25-0.19, -\y-0.5)
                +(0, 0.6) -- +(0, -0.6)
            ;
            \draw [{no-y!![\r]!80!black}]
                (2, -\y-0.5) node {$\rho_\y = \r$};
        }
        \draw
            (0, -1) node {$g^1 = 0\,0\,0\,0$}
            (0, -2) node {$g^2 = 1\,0\,0\,0$}
            (0, -3) node {$g^3 = 1\,1\,0\,0$}
            (0, -4) node {$g^4 = 0\,1\,0\,0$}
            (0, -5) node {$g^5 = 0\,1\,1\,0$}
            (0, -6) node {$g^6 = 1\,1\,1\,0$}
            (0, -7) node {$g^7 = 1\,0\,1\,0$}
            (0, -8) node {$g^8 = 0\,0\,1\,0$}
            (0, -9) node {$g^9 = 0\,0\,1\,1$}
        ;
    \end{tikzpicture}\]
    are the first nine strings.
    (Digits that are flipped are highlighted.)

    A \emph{robust Gray code} \cite{ALP22,LoP24}
    encodes integers as binary strings
    such that they can be fuzzily recovered
    even if some bits are erased or corrupted.
    Given the motivational Figure~\ref{fig:tape},
    let us use the binary symmetric channels (BSC)
    with crossover probability $p \in (0, 1/2)$ to model the errors.
    Then a robust Gray code is a pair of encoder
    \[ \EEE\colon [m] \to \{0, 1\}^{1\times n} \]
    and decoder
    \[ \DDD\colon \{0, 1\}^{1\times n} \to [m] \]
    such that
    (a) $\EEE(x)$ and $\EEE(x + 1)$ differ by one bit and
    (b)
    \[
        \Prob \Bigl\{
            \Bigl| \DDD \bigl( \BSC_p^n(\EEE(x)) \bigr) - x \Bigr| > t
        \Bigr\}
        < 2^{-\Omega(n)} + 2^{-\Omega(t)}                \label{eq:tail}
    \]
    for all $x \in [m - 1]$ and all $t > 1$.
    Here, $\BSC_p^n$ flips each of the $n$ bits with probability $p$.
    Note that \eqref{eq:tail} is almost as good as the Laplace mechanism
    in that $2^{-\Omega(t)}$ decays exponentially in $t$.
    The only catch is that when $t \gg n$,
    the other error term $2^{-\Omega(n)}$ dominates $2^{-\Omega(t)}$.
    This $2^{-\Omega(n)}$ is unavoidable because there is always
    a $2^{-\OOO(n)}$ chance that BSC will flip all ones to zero.

    Apart from robustness,
    we also care about space efficiency.
    We know that, by Shannon's theory,
    the code rate $\log_2(m)/n$ cannot exceed the capacity of $\BSC^p$,
    which is $1 + p \log_2(p) - (1 - p) \log_2(1 - p)$.
    But how close can they be?
    Before our work, Lolck and Pagh's construction \cite{LoP24}
    achieves $1/4$ of the capacity
    and Fathollahi and Wootters's construction \cite{FaW24}
    achieves $1/2$ of the capacity.
    This means that the latter uses half of the space
    to achieve the same privacy level.

    In this work, and in a concurrent work by
    Con, Fathollahi, Gabrys, and Yaakobi,
    we will show that the capacity can be achieved.
    This means that, subject to the framework of Figure~\ref{fig:tape},
    the tradeoff between privacy and space is now asymptotically tight.
    We also show that our code has linear encoding
    and decoding complexity, meaning that
    even the speed cannot be significantly improved.

\pgfdeclarelindenmayersystem{H}{
    \symbol{+}{\pgflsystemturnright}
    \symbol{-}{\pgflsystemturnleft}
    \rule{L -> +RF-LFL-FR+}
    \rule{R -> -LF+RFR+FL-}
}

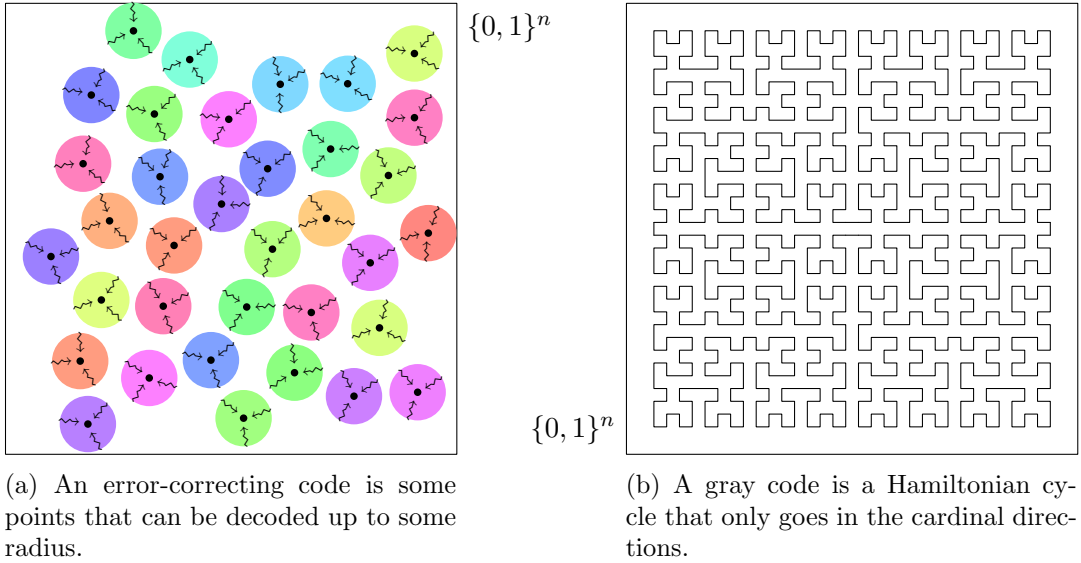
\begin{figure}
    \centering
    \pgfmathsetseed{8881616}
    \begin{subfigure}[t]{6cm}
        \begin{tikzpicture}
            \makeatletter
            \resetcolorseries[100]{no-y}
            \draw (-3, 3) rectangle (3, -3);
            \draw [overlay] (3, 3) node [below right] {$\{0, 1\}^n$};
            \pgfinterruptboundingbox
            \foreach \t in {0.5, ..., 70} {
                \pgfmathsetmacro\a{mod(\t*22.25, 36) * 10}
                \pgfmathsetmacro\c{random(0, 100)}
                \draw (\a:{sqrt(\t)*0.5}) coordinate (O);
                \pgfpointanchor{O}{center}
                \ifdim \pgf@x > -2.65cm
                \ifdim \pgf@x < 2.65cm
                \ifdim \pgf@y > -2.65cm
                \ifdim \pgf@y < 2.65cm
                \fill [{no-y!![\c]!50!white}] (O) circle (3/8);
                \fill(O) circle (0.05);
                \pgfmathsetmacro\p{rnd*360}
                \foreach \j in {1, 2, 3} {
                    \pgfmathsetmacro\a{\p + \j*120 + rnd*60}
                    \draw
                        (O) +(\a:0.25)
                        node [rotate=\a+180, scale=0.8] {$\leadsto$}
                    ;
                }
                \fi\fi\fi\fi
            }
            \endpgfinterruptboundingbox
        \end{tikzpicture}
        \caption{
            An error-correcting code is some points
            that can be decoded up to some radius.
        }
    \end{subfigure}
    \hskip1cm\hfil
    \begin{subfigure}[t]{6cm}
        \begin{tikzpicture}
            \draw (-3, 3) rectangle (3, -3);
            \draw [overlay] (-3, -3) node [above left] {$\{0, 1\}^n$};
            \tikzset{l-system={H, axiom=L, order=4, step=0.17cm}}
            \draw l-system [shift={(0.085, -0.085)}];
            \draw l-system [shift={(-0.085 - 2.55, -0.085)}];
            \draw l-system [shift={(-0.085, 0.085)}, rotate=180];
            \draw l-system [shift={(0.085 + 2.55, 0.085)}, rotate=180];
            \draw
                (-0.085 - 2.55, 0.085) -- (-0.085 - 2.55, -0.085) 
                (-0.085, 0.085) -- (0.085, 0.085) 
                (-0.085, -0.085) -- (0.085, -0.085) 
                (0.085 + 2.55, 0.085) -- (0.085 + 2.55, -0.085) 
            ;
        \end{tikzpicture}
        \caption{
            A gray code is a Hamiltonian cycle that
            only goes in the cardinal directions.
        }
    \end{subfigure}
    \caption{
        Figurative illustrations of
        error-correcting codes and Gray codes.
    }                                                   \label{fig:code}
\end{figure}

\subsection{Previous approaches}

    Earlier works \cite{LoP24, FaW24}
    baked robust Gray codes with the following recipe.
    \begin{itemize}
        \item Take a good $[n, k]$-error correcting code
            $\CCC = \{c^1, c^2, \dotsc, c^{2^k}\}
            \subset \{0, 1\}^{1\times n}$.
        \item Let $\EEE$ map ``milestone" integers
            $1 \eqqcolon \mu_1 < \mu_2 < \dotsb < \mu_{2^k} \coloneqq m$
            to the codewords of $\CCC$,
            i.e., $\EEE(\mu_j) \coloneqq c^j$.
        \item ``Interpolate'' between the milestones.
            That is, if $x \in [\mu_j, \mu_{j+1}]$,
            then the prefix of $\EEE(x)$ will come from $\EEE(\mu_j)$
            and the suffix from $\EEE(\mu_{j+1})$.
    \end{itemize}
    The technicality is with the third bullet point.
    A decoder of $\CCC$ can translate $\EEE(x)$
    back to $\mu_j$ if $x$ is close enough to $\mu_j$.
    But there is going to be a middle ground between $\mu_j$
    and $\mu_{j+1}$ such that the decoder will be confused.

    To eliminate the confusion, Lolck and Pagh \cite{LoP24}
    proposed the following data structure
    \[
        \EEE(\mu_j) \coloneqq c^j \| c^j \| c^j \| c^j
        \in \{0, 1\}^{1\times4n},
    \]
    where $\|$ is the string concatenation operator.
    They then interpolate between consecutive milestones
    $\mu_j$ and $\mu_{j+1}$ as
    \[\def\arraycolsep{0pt}
    \def\!{\tikz [overlay] \draw
        [line width=0.55cm, yellow!80!black] (0.15, 0.1) -- +(0, -0.4);}
    \begin{array}{rccccccc}
        \EEE(\mu_{j.1}) \coloneqq {}
        & \!c^j &\|& c^j &\|& c^j &\|& c^j,
        \\ \EEE(\mu_{j.2}) \coloneqq {}
        & c^{j+1} &\|& \!c^j &\|& c^j &\|& c^j,
        \\ \EEE(\mu_{j.3}) \coloneqq {}
        & c^{j+1} &\|& c^{j+1} &\|& \!c^j &\|& c^j,
        \\ \EEE(\mu_{j.4}) \coloneqq {}
        & c^{j+1} &\|& c^{j+1} &\|& c^{j+1} &\|& \!c^j,
        \\ \EEE(\mu_{j.5}) \coloneqq {}
        & c^{j+1} &\|& c^{j+1} &\|& c^{j+1} &\|& c^{j+1}
    \end{array}\]
    for some minor milestones $\mu_j \eqqcolon \mu_{j.1} < \mu_{j.2}
    < \mu_{j.3} < \mu_{j.4} < \mu_{j.5} \coloneqq \mu_{j+1}$.
    Note that only one copy is undergoing interpolation
    at any given time (which is highlighted).
    So the advantage of repeating $c^j$ four times is that
    there are always two other copies that will decode
    to the same codeword.
    To elaborate, between $\mu_{j.1}$ and $\mu_{j.3}$,
    the two $c^j$ to the right will decode correctly;
    between $\mu_{j.3}$ and $\mu_{j.5}$, the two $c^{j+1}$
    to the left will decode correctly.
    
    Later, Fathollahi and Wootters \cite{FaW24}
    streamlined the data structure from $4n$ bits to $(2 + 3\ee) n$ bits
    by using \emph{buffers}---consecutive zeros and ones.
    They map milestones to
    \[
        \EEE(\mu_j) \coloneqq
        0^{\ee n} \| c^j \| 0^{\ee n} \| c^j \| 0^{\ee n}
        \in \{0, 1\}^{1\times(2+3\ee)n}
    \]
    if $j$ is even, and to
    \[
        \EEE(\mu_j) \coloneqq
        1^{\ee n} \| c^j \| 1^{\ee n} \| c^j \| 1^{\ee n} 
        \in \{0, 1\}^{1\times(2+3\ee)n}
    \]
    if $j$ is odd.
    They then interpolate between the milestones as
    \[\def\arraycolsep{0pt}
    \def\!{\tikz [overlay] \draw
        [line width=0.55cm, yellow!80!black] (0.25, 0.1) -- +(0, -0.4);}
    \begin{array}{rccccccccc}
        \EEE(\mu_{j.1}) \coloneqq {} & \!0^{\ee n} &\|& c^j
        &\|& 0^{\ee n} &\|& c^j &\|& 0^{\ee n},
        \\ \EEE(\mu_{j.2}) \coloneqq {} & 1^{\ee n} &\|& \!c^j
        &\|& 0^{\ee n} &\|& c^j &\|& 0^{\ee n},
        \\ \EEE(\mu_{j.3}) \coloneqq {} & 1^{\ee n} &\|& c^{j+1}
        &\|& \!0^{\ee n} &\|& c^j &\|& 0^{\ee n},
        \\ \EEE(\mu_{j.4}) \coloneqq {} & 1^{\ee n} &\|& c^{j+1}
        &\|& 1^{\ee n} &\|& \!c^j &\|& 0^{\ee n},
        \\ \EEE(\mu_{j.5}) \coloneqq {} & 1^{\ee n} &\|& c^{j+1}
        &\|& 1^{\ee n} &\|& c^{j+1} &\|& \!0^{\ee n},
        \\ \EEE(\mu_{j.6}) \coloneqq {} & 1^{\ee n} &\|& c^{j+1}
        &\|& 1^{\ee n} &\|& c^{j+1} &\|& 1^{\ee n}
    \end{array}\]
    for some minor milestones $\mu_j \eqqcolon \mu_{j.1} < \mu_{j.2} <
    \mu_{j.3} < \mu_{j.4} < \mu_{j.5} < \mu_{j.6} \coloneqq \mu_{j+1}$.
    In this construction,
    the decoder is left with two, not four, copies of $c^j$.
    It knows that the one sandwiched between $0^{\ee n}$ and $1^{\ee n}$
    is the one undergoing interpolation,
    and hence the other one will decode correctly.
    To be more precise, between $\mu_{j.1}$ and $\mu_{j.4}$,
    the left one is undergoing interpolation
    and the right $c^j$ is trustworthy;
    between $\mu_{j.3}$ and $\mu_{j.6}$,
    the right one is undergoing interpolation
    and the left $c^{j+1}$ is trustworthy.

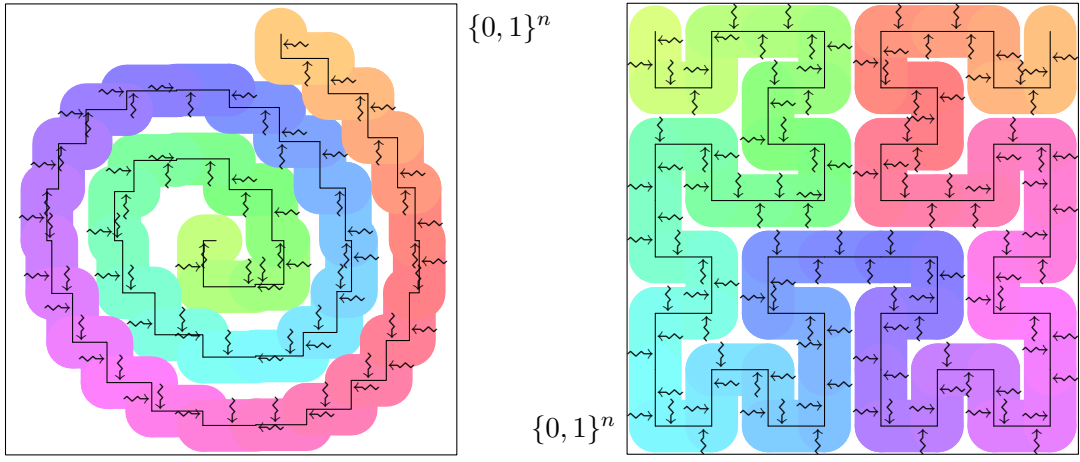
\begin{figure}
    \centering
    \pgfmathsetseed{8881616}
    \begin{subfigure}[t]{6cm}
        \begin{tikzpicture}
            \resetcolorseries[41]{no-y}
            \draw (-3, 3) rectangle (3, -3);
            \draw [overlay] (3, 3) node [below right] {$\{0, 1\}^n$};
            \tikzset{shift={(-0.2, -0.15)}}
            \xdef\a{0}
            \xdef\b{0}
            \draw (0, 0) coordinate (B);
            \foreach \t in {2, ..., 41} {
                \draw (B) coordinate (A);
                \draw ({sqrt(\t)*180}:{sqrt(\t)*0.45}) coordinate (B);
                \draw [{no-y!![\t]!50!white}, line width=0.7cm]
                    (A) -| (B);
            }
            \def\rot#1{\rotatebox{#190}{$\leadsto$}}
            \xdef\a{0}
            \xdef\b{0}
            \draw (0, 0) coordinate (B);
            \foreach \t in {2, ..., 41} {
                \draw (B) coordinate (A);
                \draw ({sqrt(\t)*180}:{sqrt(\t)*0.45}) coordinate (B);
                \draw [every node/.style={
                        sloped, allow upside down, auto, inner sep=0pt
                    }]
                    (A) -|
                    node [pos=0.25+rand*0.2] {\rot-}
                    node [pos=0.75+rand*0.2, '] {\rot+}
                    (B)
                ;
            }
        \end{tikzpicture}
        \caption{
            ``Grayed code'': Old approach takes an error correcting code
            and then interpolates between codewords.
        }
    \end{subfigure}
    \hskip1cm\hfil
    \begin{subfigure}[t]{6cm}
        \begin{tikzpicture}
            \resetcolorseries[64]{no-y}
            \draw (-3, 3) rectangle (3, -3);
            \draw [overlay] (-3, -3) node [above left] {$\{0, 1\}^n$};
            \xdef\c{0}
            \tikzset{shift={(-2.625, 2.625)}}
            \tikzset{l-system={H, axiom=L, order=3, step=0.75cm}}
            \draw l-system
                [decorate, decoration={
                    show path construction,
                    lineto code={
                        \xdef\c{\the\numexpr\c+1}
                        \draw [{no-y!![\c]!50!white}, line width=0.7cm]
                            (\tikzinputsegmentfirst) --
                            (\tikzinputsegmentlast)
                        ;
                    },
                }]
            ;
            \def\rot#1{\rotatebox{#190}{$\leadsto$}}
            \draw l-system
                [decorate, decoration={
                    show path construction,
                    lineto code={
                        \draw [every node/.style={
                                sloped, allow upside down, auto,
                                inner sep=0pt
                            }]
                            (\tikzinputsegmentfirst) --
                            node [pos=0.25+rand*0.2] {\rot-}
                            node [pos=0.75+rand*0.2, '] {\rot+}
                            (\tikzinputsegmentlast)
                        ;
                    },
                }]
            ;
        \end{tikzpicture}
        \caption{
            ``Coded Gray code'': New approach multiplies a Gray code
            with the generator matrix of an error-correcting code.
        }
    \end{subfigure}
    \caption{
        Old approach (not capacity-achieving) versus
        new approach (capacity-achieving).
    }                                                \label{fig:combine}
\end{figure}

\subsection{New approach}

    \begin{tikzpicture} [overlay, shift={(-0.5, 0.3)}, yscale=0.02]
        \draw [line width=0.5cm, yellow!80!black] (0, -1) -- (0, -64);
    \end{tikzpicture}%
    \itshape
    While this paper was in preparation,
    it came to our attention that
    Con, Fathollahi, Gabrys, \hbox{Wootters}, and Yaakobi
    have achieved similar results, but with different techniques
    \cite{CFG24}.
    In particular, their approach uses code concatenation.
    \upshape
    \bigskip

    In this and the concurrent work
    by Con, Fathollahi, Gabrys, Wootters, and Yaakobi,
    we aim to rightsize the length to $n + \Theta(\ee n)$ bits.
    While their work uses code concatenation,
    we begin with a generator matrix $A \in \{0, 1\}^{k\times n}$
    of some error-correcting code.
    We then reorder the codewords $c^1, \dotsc, c^{2^k}$
    using Gray code:
    \[
        c^{j+1} = g^{j+1} A = (g^j + e^{\rho_j}) A = c^j + A^{\rho_j}
        \in \{0, 1\}^{1\times k}.
    \]
    Here, $g^j$ is the $j$th string of the Gary code,
    $e^{\rho_j}$ is the $\rho_j$th cardinal vector, and
    $A^{\rho_j}$ is the $\rho_j$th row of $A$, all as row vectors.
    Our data structure will look like
    \[
        c^j \| 0^{\ee n} \| \rho_j \| \beta^j \| 0^{\ee n}
        \| \rho_j \| \beta^j \| 0^{\ee n}
    \]
    or
    \[
        c^j \| 1^{\ee n} \| \rho_j \| \beta^j \| 1^{\ee n}
        \| \rho_j \| \beta^j \| 1^{\ee n}
    \]
    depending on the parity of $j$,
    Here, $\beta^j$ is a subvector of $c^j$ obtained by
    collecting bits where $A^{\rho_j}$ has $1$.
    More precisely, if $A^{\rho_j}$ has $1$ at indices
    $i_1, i_2, \dotsc, i_w$, then
    $\beta^j \coloneqq c^j_{i_1} c^j_{i_2} \dotsm c^j_{i_w}
    \in \{0, 1\}^{1\times w}$,
    where $w$ is the Hamming weight of $A^{\rho_j}$.
    
    The purpose of keeping $\rho_j$ in $\EEE$
    is to take note of which row of $A$
    we are going to add to $c^j$ to obtain $c^{j+1}$.
    The purpose of keeping $\beta^j$ in $\EEE$
    is to back up the bits of $c^j$ that are going to be modified.
    We then interpolate between minor milestones
    \[\def\arraycolsep{0pt}
    \def\!{\tikz [overlay] \draw
        [line width=0.55cm, yellow!80!black] (0.25, 0.1) -- +(0, -0.4);}
    \begin{array}{ccccccccccccccc}
        \!c^j &\|& 0^{\ee n} &\|& \rho_j &\|& \beta^j
        &\|& 0^{\ee n} &\|& \rho_j &\|& \beta^j &\|& 0^{\ee n},
        \\ c^{j+1} &\|& \!0^{\ee n} &\|& \rho_j &\|& \beta^j
        &\|& 0^{\ee n} &\|& \rho_j &\|& \beta^j &\|& 0^{\ee n},
        \\ c^{j+1} &\|& 1^{\ee n} &\|& \!\rho_j &\|& \beta^j
        &\|& 0^{\ee n} &\|& \rho_j &\|& \beta^j &\|& 0^{\ee n},
        \\ c^{j+1} &\|& 1^{\ee n} &\|& \rho_{j+1} &\|& \!\beta^j
        &\|& 0^{\ee n} &\|& \rho_j &\|& \beta^j &\|& 0^{\ee n},
        \\ c^{j+1} &\|& 1^{\ee n} &\|& \rho_{j+1} &\|& \beta^{j+1}
        &\|& \!0^{\ee n} &\|& \rho_j &\|& \beta^j &\|& 0^{\ee n},
        \\ c^{j+1} &\|& 1^{\ee n} &\|& \rho_{j+1} &\|& \beta^{j+1}
        &\|& 1^{\ee n} &\|& \!\rho_j &\|& \beta^j &\|& 0^{\ee n},
        \\ c^{j+1} &\|& 1^{\ee n} &\|& \rho_{j+1} &\|& \beta^{j+1}
        &\|& 1^{\ee n} &\|& \rho_{j+1} &\|& \!\beta^j &\|& 0^{\ee n},
        \\ c^{j+1} &\|& 1^{\ee n} &\|& \rho_{j+1} &\|& \beta^{j+1}
        &\|& 1^{\ee n} &\|& \rho_{j+1} &\|& \beta^{j+1} &\|& \!0^{\ee n},
        \\ c^{j+1} &\|& 1^{\ee n} &\|& \rho_{j+1} &\|& \beta^{j+1}
        &\|& 1^{\ee n} &\|& \rho_{j+1} &\|& \beta^{j+1} &\|& 1^{\ee n},
    \end{array}\]
    Note that we use Fathollahi and Wootters's data structure
    to protect $\rho_j$ and $\beta^j$, and so the backup data
    is almost always available.\footnote{
    There is no reason not to protect $\rho_j$ and $\beta^j$
    using error-correcting codes.
    We omit that here but will discuss in the formal proof.}

    An analogy of this construction is to think of
    $c_j$ as the state of our computer at version $j$.
    Now a software update comes in and attempts to
    add $A^{\rho_j}$ to $c_j$.
    To avoid messing things up, the updater backs up the files
    that are going to be updated, which are at $i_1, \dotsc, i_w$;
    and $\beta^j$ is the backup data.
    
    Our construction, at any code rate below capacity,
    achieves positive error exponents
    and linear encoding and decoding complexity.

    \begin{theorem} [Main theorem]                      \label{thm:main}
        Fix a BSC with $p \in (0, 1/2)$
        and a gap to capacity $\ee > 0$.
        For sufficiently large $n$,
        there exists a pair of
        encoder $\EEE\colon [m] \to \{0, 1\}^{1\times n}$ and
        decoder $\DDD\colon \{0, 1\}^{1\times n} \to [m]$
        with code size $m > 2^{(\Capacity(\BSC_p) - \ee) n}$
        such that
        (a) $\EEE(x)$ and $\EEE(x + 1)$ differ by one bit and
        (b)
        \[
            \Prob \Bigl\{
                \Bigl| \DDD \bigl( \BSC_p^n(\EEE(x)) \bigr) - x \Bigr|
                > t
            \Bigr\}
            < 2^{-\Omega(n)} + 2^{-\Omega(t)}
        \]
        for all $x \in [m - 1]$ and all $t > 1$.
        Moreover, the complexity of $\EEE$ and $\DDD$
        scales\footnote{exponentially in $1/\poly(\ee)$}
        linearly in $n$.
    \end{theorem}

    \paragraph{Organization:}
    The rest of the paper is dedicated to
    proving Theorem~\ref{thm:main}.

\section{Proof of Theorem~\ref{thm:main}}

    Fix a crossover probability $p$
    and a gap to capacity $\ee > 0$.
    Let $n$ and $k$ be very large.

\subsection{The building blocks
              \texorpdfstring{$\BBB$}{B} and \texorpdfstring{$\CCC$}{C}}

    We begin with a linear code $\BBB$
    with block length $\ee n$
    and dimension $\ee k$.
    Using well-known constructions \cite[Theorem~8]{GuI05},
    we can make the code rate $k/n$ $\ee$-close to the capacity
    if $n$ is large enough.
    Moreover, the encoding and decoding complexity
    can be made linear in $n$.
    Let $B$ be the generator matrix of $\BBB$.

    We stack $B$ to construct a larger generator matrix
    \[
        A \coloneqq \bma{
            B & \bar B
            \\ & B & \bar B
            \\ && B & \bar B
            \\ &&& \ddots & \ddots
            \\ &&&& B & \bar B
        }
        \in \{0, 1\}^{k\times(1+\ee)n}                      \label{eq:A}
    \]
    and denote the corresponding code by
    $\CCC \subset \{0, 1\}^{1\times(1+\ee)n}$.
    Here, $\bar B$ is the bitwise complement of $B$.
    We put $\bar B$ next to $B$ so that all rows of $[B \quad \bar B]$
    has the same Hamming weight, $\ee n$.
    This means that the backup data $\beta^j$
    will be exactly $\ee n$ bits long.
    We repeat $B$ and $\bar B$ $1/\ee$ times so that
    $A$ has block length $(1 + \ee) n$ and dimension $k$.
    This makes the code rate of $\CCC$ $2\ee$-close to the capacity.
    
    Decoding $\CCC$ is straightforward.
    Given a received word $y \in \{0, 1\}^{1\times(1+\ee)n}$,
    apply $\BBB$'s decoder to $y_1, \dotsc, y_{\ee n}$
    to obtain $x_1, \dotsc, x_{\ee n}$.
    Subtract the influence of $x_1, \dotsc, x_{\ee n}$
    from $y_{\ee n+1}, \dotsc, y_{2\ee n}$
    and apply $\BBB$'s decoder to obtain
    $x_{\ee n+1}, \dotsc, x_{2\ee n}$.
    Repeat this process until we obtain $x_n$.

    We also use $\BBB$ to protect the row index $\rho_j \in [k]$
    and the backup data $\beta^j \in \{0, 1\}^{\ee n}$.
    Denote by $\BBB(\rho_j, \beta^j)$ the result of
    encoding these $\log_2(k) + \ee n$ bits of information
    using
    \[
        \frac{2n}{k} \geq
        \Bigl\lceil
            \frac{\log_2(k) + \ee n}{\ee k}
        \Bigr\rceil
    \]
    blocks of $\BBB$.
    This means that $\BBB(\rho_j, \beta^j)$ has length $2 \ee n^2 / k$.

\subsection{Encoding \texorpdfstring{$\EEE$}{}}

    Recall that $\rho_j$ is the ruler sequence
    defined in \eqref{eq:ruler} capped at $k$.
    Recall that $g^j$ is the $j$th string of the Gray code and is
    obtained by flipping the $\min(\rho_{j-1}, k)$th bit of $g^{j-1}$.
    We assume an ordering on the codewords
    $\CCC = \{c^1, \dotsc, c^{2^k}\}$ by Gray code, i.e.,
    $c^j \coloneqq g^j A$.
    Let $\beta^j$ be the subvector of $c^j$
    obtained by deleting the bits where $A^{\rho_j}$ has $0$.

    We now place the milestones at
    \[ \mu_j \coloneqq j \ee n (4 + 2n/k) \]
    for $j \in [2^k]$.
    Consequently, $\EEE$ will encode integers up to
    $m \coloneqq (2^k - 1) \ee n (4 + 2n/k) + 1 = (1 + o(1)) 2^k$.
    We then define data structure:
    \[
        \EEE(\mu_j) \coloneqq c^j \| 0^{\ee n} \| \BBB(\rho_j, \beta^j)
        \| 0^{\ee n} \| \BBB(\rho_j, \beta^j) \| 0^{\ee n}
    \]
    and
    \[
        \EEE(\mu_j) \coloneqq c^j \| 1^{\ee n} \| \BBB(\rho_j, \beta^j)
        \| 1^{\ee n} \| \overline{\BBB(\rho_j, \beta^j)} \| 1^{\ee n}
    \]
    depending on the parity of $j$.
    Here, $\overline{\BBB(\rho_j, \beta^j)}$
    is the bitwise complement of $\BBB(\rho_j, \beta^j)$.
    Note that each $\EEE(\mu_j)$
    is $(1 + 4\ee + 4\ee n/k) n$ bits long.
    We infer that the code rate of $\EEE$
    is $\OOO(\ee)$-close to the capacity.

    Next, we show that the Hamming distance between
    $\EEE(\mu_j)$ and $\EEE(\mu_{j+1})$ is $\ee n (4 + 2n/k)$.
    Trivially, the consecutive zeros and ones contributes $3\ee n$
    bits of Hamming distance.
    Next, note that
    \[
        c^{j+1} - c^j
        = g^{j+1} A - g^j A
        = (g^{j+1} - g^j) A
        = e^{\rho_j} A
        = A^{\rho_j}.
    \]
    This is the $\rho_j$th row of $A$.
    By the construction \eqref{eq:A}, any row of $A$
    contributes exactly $\ee n$ bits of Hamming distance.
    Next, $\BBB(\rho_j, \beta^j)$ and $\BBB(\rho_{j+1}, \beta^{j+1})$
    contributes an unknown amount of distance.
    But it is complement to the distance between $\BBB(\rho_j, \beta^j)$
    and $\overline{\BBB(\rho_{j+1}, \beta^{j+1})}$.
    Therefore, the $\BBB$ part contributes exactly $2 \ee n^2 / k$.
    In total, the Hamming distance is exactly $\ee n (4 + 2n/k)$.

    Now that the distance between consecutive milestones
    matches the Hamming distance,
    we can interpolate between them.
    In particular, for even $j$,
    \[\def\arraycolsep{0pt}
    \def\!{\tikz [overlay] \draw
        [line width=0.55cm, yellow!80!black] (0.25, 0.1) -- +(0, -0.4);}
    \def\?{\tikz [overlay] \draw
        [line width=0.55cm, yellow!80!black] (0.8, 0.1) -- +(0, -0.4);}
    \begin{array}{rccccccccccc}
        \EEE(\mu_j) \coloneqq {} & \!c^j
        &\|& 0^{\ee n} &\|& \BBB(\rho_j, \beta^j)
        &\|& 0^{\ee n} &\|& \BBB(\rho_j, \beta^j) &\|& 0^{\ee n},
        \\ \EEE(\mu_j + \ee n) \coloneqq {} & c^{j+1}
        &\|& \!0^{\ee n} &\|& \BBB(\rho_j, \beta^j)
        &\|& 0^{\ee n} &\|& \BBB(\rho_j, \beta^j) &\|& 0^{\ee n},
        \\ \EEE(\mu_j + 2\ee n) \coloneqq {} & c^{j+1}
        &\|& 1^{\ee n} &\|& \?\BBB(\rho_j, \beta^j)
        &\|& 0^{\ee n} &\|& \BBB(\rho_j, \beta^j) &\|& 0^{\ee n},
        \\ \EEE(\mu_j + 2\ee n + d) \coloneqq {} & c^{j+1}
        &\|& 1^{\ee n} &\|& \BBB(\rho_{j+1}, \beta^{j+1})
        &\|& \!0^{\ee n} &\|& \BBB(\rho_j, \beta^j) &\|& 0^{\ee n},
        \\ \EEE(\mu_j + 3\ee n + d) \coloneqq {} & c^{j+1}
        &\|& 1^{\ee n} &\|& \BBB(\rho_{j+1}, \beta^{j+1})
        &\|& 1^{\ee n} &\|& \?\BBB(\rho_j, \beta^j) &\|& 0^{\ee n},
        \\ \EEE(\mu_j + 3\ee n + 2 \ee n^2 / k) \coloneqq {} & c^{j+1}
        &\|& 1^{\ee n} &\|& \BBB(\rho_{j+1}, \beta^{j+1})
        &\|& 1^{\ee n} &\|& \overline{\BBB(\rho_{j+1}, \beta^{j+1})}
        &\|& \!0^{\ee n},
        \\ \EEE(\mu_j + 4\ee n + 2 \ee n^2 / k) \coloneqq {} & c^{j+1}
        &\|& 1^{\ee n} &\|& \BBB(\rho_{j+1}, \beta^{j+1})
        &\|& 1^{\ee n} &\|& \overline{\BBB(\rho_{j+1}, \beta^{j+1})}
        &\|& 1^{\ee n},
    \end{array}\]
    where $d$ is the Hamming distance between
    $\BBB(\rho_j, \beta^j)$ and $\BBB(\rho_{j+1}, \beta^{j+1})$.

\subsection{Decoding \texorpdfstring{$\DDD$}{D}}

    Suppose that we are given
    \[
        c \| \phi \| B' \| \phi' \| B'' \| \phi'''      \label{eq:noisy}
    \]
    as the noisy version of $\EEE(x)$ for some $x \in [m]$,
    where
    \begin{itemize}
        \item $c \in \{0, 1\}^{1\times(1+\ee)n}$
            is the noisy version of $c^j$, $c^{j+1}$,
            or anything in between,
        \item $\phi, \phi', \phi'' \in \{0, 1\}^{1\times\ee n}$
            are the noisy version of the buffers, and
        \item $B', B'' \in \{0, 1\}^{1\times2\ee n^2/k}$
            are the noisy version of the $\BBB$ part.
    \end{itemize}
    We first apply Fathollahi and Wootters's decoder \cite{FaW24}
    to the second half of \eqref{eq:noisy}
    \[ \phi \| B' \| \phi' \| B'' \| \phi'''. \]
    Their decoder counts how many ones and zeros are
    in $\phi$, $\phi'$, and $\phi''$.
    This tells us which minor milestone we are at.
    We use this information to determine which of $B'$ or $B''$
    is undergoing interpolation, and which is trustworthy.
    And then, we use the trustworthy one to recover
    the row index $\rho_j$ and the backup data $\beta^j$
    (or $\rho_{j+1}$ and $\beta^{j+1}$ depending on
    if $x$ is past $\mu_j + 2.5 \ee n + d$ or not).
    From now on, we just call them $\rho$ and $\beta$.

    We define the rollback function
    $\Roll\colon \{0, 1\}^{(1+\ee)n} \times [k] \times \{0, 1\}^{\ee n}
    \to \{0, 1\}^{(1+\ee)n}$
    that overwrites messed-up bits using backup data.
    More precisely, $\Roll(c, \rho, \beta)$ will be the vector $c$
    after replacing $c_{i_1}$ with $\beta_1$,
    $c_{i_2}$ with $\beta_2$, and so on, where $i_1, i_2, \dotsc$
    are the indices where $A^\rho$ has $1$.
    Our claim is that, it does not matter if it is
    $c^j$, $0^{\ee n}$, or $\BBB$ that is undergoing interpolation,
    $\Roll(c, \rho, \beta)$
    will just look like the noisy version of $c^j$ or $c^{j+1}$,
    which can be decoded by the decoder of $\CCC$.
    This can be seen more clearly
    by considering three cases.

    Case 1: $x$ is between $\mu_j$ and $\mu_j + \ee n$.
    This is the stage where we are adding $A^\rho$ to $c^j$.
    In this case, the $\BBB$ part backs up the subvector of $c^j$
    that is undergoing interpolation;
    $\Roll(c, \rho, \beta)$ would just be a noisy version of $c^j$
    that can be decoded by $\CCC$.

    Case 2: $x$ is between $\mu_j + \ee n$ and $\mu_j + 2.5 \ee n + d$.
    This is the case where $B'$ is not trustworthy
    and so Fathollahi and Wootters's decoder
    will decode $B''$ to $(\rho_j, \beta^j)$.
    In this case, $\Roll(c, \rho_j, \beta^j)$ will be
    a noisy version of $c^j$ that can be decoded by $\CCC$.
    
    Case 3: $x$ is between $\mu_j + 2.5 \ee n + d$
    and $\mu_j + 4\ee n + 2 \ee n^2 / k$.
    This is the case where $B''$
    is not trustworthy and so Fathollahi and Wootters's decoder
    will decode $B'$ to $\rho_{j+1}, \beta^{j+1}$.
    In this case, $\Roll(c, \rho_{j+1}, \beta^{j+1})$ will be
    a noisy version of $c^{j+1}$ that can be decoded by $\CCC$.

    Examining these three cases, we can see that
    $\Roll(c, \rho, \beta)$
    will always yield $c^j$ or $c^{j+1}$.
    With that, we can compute $g^j$ and $j$.
    Now that we know $x \in [\mu_j, \mu_{j+1}]$,
    it suffices to compare \eqref{eq:noisy}
    with $\EEE(\mu_j), \dotsc, \EEE(\mu_{j+1})$
    and see which one minimizes the Hamming distance.
    The minimizer will be our best bet of $x$.

\subsection{Complexity and tail estimation}

    The complexity of $\EEE$ and $\DDD$ is linear in $n$.
    This is because Gray's encoding, Gray's decoding,
    $\BBB$'s encoding, $\BBB$'s decoding,
    determining whether $\phi$, $\phi'$, and $\phi''$ are zeros or ones,
    determining whether $B'$ or $B''$ is trustworthy,
    and $\Roll$ are all linear in $n$.

    The tail estimation \eqref{eq:tail}
    boils down to the following components.
    \begin{itemize}
        \item With probability $2^{-\Omega(n)}$,
            we obtain the wrong $j$, i.e.,
            $x \notin [\mu_j, \mu_{j+1}]$.
        \item The guesswork of $x$ conditioned on correct $j$
            has tail probability $2^{-\Omega(t)}$.
    \end{itemize}
    The first bullet point is a consequence of
    the error probability of $\BBB$ being $2^{-\Omega(\ee n)}$,
    which is $2^{-\Omega(n)}$ as we fixed $\ee$.
    The second bullet point relies on
    what minimizes the Hamming distance between \eqref{eq:noisy}
    and $\EEE(\mu_j), \dotsc, \EEE(\mu_{j+1})$.
    Such analysis has been done before
    \cite[Lemma~3.7]{LoP24} \cite[Lemma~13]{FaW24},
    and we do not repeat it here.
    This finishes the proof.

\bibliographystyle{alpha}
\bibliography{ShannonGray-51}

\end{document}